\makeatletter
\declare@file@substitution{revtex4-1.cls}{revtex4-2.cls}
\makeatother

\documentclass[trackchanges]{aastex63} 
\usepackage{graphicx}
\usepackage{natbib}
\usepackage{amsmath,bm}
\usepackage{lineno}

\begin{document} 
\title{Ion-Scale Waves Regulated by Plasma Beta and Cross Helicity: Observational Evidence for the Helicity Barrier in Imbalanced Solar Wind Turbulence}

%
\author{G. Q. Zhao}
\affiliation{Institute of Space Physics, Luoyang Normal University, Luoyang, China}
\affiliation{Henan Key Laboratory of Electromagnetic Transformation and Detection, Luoyang, China}

\author{R. Meyrand}  
\affiliation{Space Science Center and Department of Physics and Astronomy, University of New Hampshire, Durham, USA}
\affiliation{Department of Physics, University of Otago, Dunedin, New Zealand}

\author{H. Q. Feng} 
\affiliation{Institute of Space Physics, Luoyang Normal University, Luoyang, China}

\author{H. F. Yang} 
\affiliation{Institute of Space Physics, Luoyang Normal University, Luoyang, China}
\affiliation{Research Center for Novel Solar-Blind Ultraviolet and Infrared Photoelectronic Detectors of Henan province, Luoyang, China}

\author{L. Xiang} 
\affiliation{Institute of Space Physics, Luoyang Normal University, Luoyang, China}

\author{Z. Wang} 
\affiliation{Institute of Space Physics, Luoyang Normal University, Luoyang, China}

\begin{abstract}
Ion-scale waves are believed to play an important role in heating of the solar corona and wind, though their generation mechanism remains unclear. Based on Parker Solar Probe observations, this Letter investigates the occurrence of ion-scale waves in the near-Sun solar wind with heliocentric distances between 0.1 and 0.2 au. Results show that the occurrence rate of left-handed polarized waves significantly depends on the plasma beta ($\beta$) and cross helicity ($\sigma_c$). The occurrence rate rapidly increases with decreasing $\beta$ and increasing $\sigma_c$. Overall, the occurrence rate exceeds 45$\%$ when $\beta<0.2$ and $\sigma_c > 0.7$ are satisfied. These observations are consistent with the direct predictions of the helicity barrier theory that suggests the generation of ion cyclotron waves via imbalanced magnetized turbulence. 

\end{abstract}


\section{Introduction}
Ion cyclotron waves (ICWs) are left-handed polarized electromagnetic waves with frequencies near the proton cyclotron frequency \citep[e.g.,][]{sti62}. They have attracted much attention due to their ability in energizing ions and/or modifying velocity distributions for a collisionless plasma \citep{hol75p63,mar06p01}. Early studies showed that the solar corona and wind suffer from heating preferentially in the direction perpendicular to the ambient magnetic field \citep{mar82p52,gaz82p43,cra02p29}. ICWs were proposed for the perpendicular heating \citep{mar82p30,cra00p97,hol02p47,ise11p88}. Observationally ion-scale waves are common in the solar wind \citep{jia14p23,zha18p15,bow20p66,hua20p03,liu23p69,liu25p80}. Their occurrence rate is particularly high up to 30$-$$50\%$ for intervals with radial magnetic fields in the near-Sun regions. Most of the waves are left-handed polarized with respect to the ambient field and are believed to be ICWs propagating away from the Sun \citep{jia09p05,bow20p74,nir26p04}. Direct measures of ion temperatures or velocity distributions in the solar wind are consistent with cyclotron resonant heating by ICWs \citep{kas08p03,zha20p14,bow22p01}. Unstable ion velocity distributions were also suggested to drive the waves via instability \citep{bal09p01,gar16p30,zha19p75,liu25p35}. Nevertheless, it is still an open question how ion-scale waves are generated in the corona as well as in the solar wind.    

A relevant mechanism termed as helicity barrier was proposed for imbalanced magnetized turbulence like the solar wind \citep{mey21p01,squ22p15}. Using the model of finite-Larmor-radius magnetohydrodynamics, \citet{mey21p01} argued that there exists a barrier at scales near the ion gyroradius ($\rho_i$) due to two conserved quantities of energy and generalized helicity. The generalized helicity reduces to be magnetohydrodynamic cross helicity ($\sigma_c$) and undergoes a forward cascade at scales above $\rho_i$, while it becomes magnetic helicity that results in an inverse cascade at scales below $\rho_i$. The collision of the two cascades generates a helicity barrier inevitably in the imbalanced turbulence. Based on six-dimensional simulations of imbalanced turbulence, \citet{squ22p15} further showed that the helicity barrier causes the energy at scales above $\rho_i$ to grow as the turbulence evolves and generates ICWs. \citet{zha25p21} conducted a multi-ion simulation and found that the helicity barrier leads to turbulent energy to accumulate at the inertial scales until oblique fluctuations from the critically balanced cascade attain ion-cyclotron frequencies. Clear signatures of oblique cyclotron resonance manifest in highly asymmetric ion velocity distributions in the simulation, suggesting ion heating by oblique ICWs. The authors also found that the proton velocity distribution ultimately becomes sufficiently unstable and drives parallel ICWs. \citet{yer26p92} provided an analytical description on the transfer of energy from oblique to parallel ICWs and argued that strong proton heating by oblique ICWs will drive parallel ICWs via a process of quasi-linear focusing \citep{cha10p10,ise11p88}. 

Two parameters are crucial for the helicity barrier mechanism. One parameter is the $\beta$ that is the ratio of thermal to magnetic pressure. The $\beta$ is required to be small so that ions cannot couple efficiently to Alfv\'enic fluctuations. The other parameter is the $\sigma_c$ that describes the level of imbalance of turbulence. The higher $\sigma_c$ means the more imbalanced turbulence. As the turbulence becomes increasingly imbalanced, a progressively smaller fraction of the turbulent energy flux is able to pass through the helicity barrier, causing energy to accumulate above ion scales and parallel ICWs to be emitted \citep{squ22p15,zha25p21,yer26p92}. In this case larger energy with scales above $\rho_i$ and more ICWs are speculated. A mild correlation (with a correlation coefficient of 0.44) between $\sigma_c$ and magnetic energy density at scales above $\rho_i$ was revealed, which was interpreted as a signature of the helicity barrier \citep{zha22p24}. Note that compelling evidence for the barrier should be the rise of significant dependence of ICW occurrence on the parameters of $\beta$ and $\sigma_c$, since this mechanism is expected to become more robust with decreasing $\beta$ and increasing $\sigma_c$ \citep{mey21p01,squ22p15}. Such dependence, to the best of our knowledge, has not been reported, although evidence for the barrier was provided with the research on the turbulence transition range and cross helicity cascade rates in the solar wind  \citep{mci25p08,pan25p60}. 

Based on Parker Solar Probe (PSP) observations, this Letter reports that the occurrence of left-handed polarized ion-scale waves can be well regulated by $\beta$ and $\sigma_c$. The observed highest occurrence rate of the waves is maximized precisely in the low $\beta$, high $\sigma_c$ regime, which is the core prediction of the helicity barrier theory. The data and analysis methods are described in Section 2. Statistical results are presented in Section 3. Section 4 is the summary and discussion. 

\section{Data and Analysis Methods}  %
The data used in this Letter are from the PSP observations during Encounters 3$-24$ with high cadences. The magnetic field ${\bm B}$ is measured by the FIELDS flux-gate magnetometer with a cadence range from 73.24 to 292.97 Hz \citep{bal16p49}. The plasma data, including proton bulk velocity ${\bm V}_p$, density $N_p$ and temperature $T_{p}$, are derived by the SPAN-I instrument \citep{liv22p38}. The plasma beta is defined by $\beta=2\mu_0N_pkT_{p}/B^2$, where $\mu_0$ is the permeability of free space, $k$ is the Boltzmann constant, $N_p$ and $T_{p}$ are from the SPAN-I level-3 moment data, and $T_{p}$ refers to the isotropic proton temperature. The cross helicity is a measure of the imbalance of counter-propagating waves. It is calculated directly via the equation $\sigma_c=2\langle{\bm v}{\cdot}{\bm b}\rangle/(\langle{\bm v}^2\rangle+\langle{\bm b}^2\rangle)$, where  ${\bm v}$ and ${\bm b}$ are the velocity and magnetic field fluctuations in velocity units, respectively, and $\langle\rangle$ represents the ensemble average. A period of 10 minutes is used to produce one value of $\sigma_c$, being at inertial scales. 

Some operations are performed to select the data as follows. First, we discard the data with $N_p < 100$ cm$^{-3}$ since the low density of protons is possibly attributed to the incomplete measurements of proton velocity distributions by SPAN-I \citep{liv22p38}. Second, we follow the approach described by \citet{rom24p00} and calculate the field-of-view (FOV) coverage parameter $A_{\text{\upshape FOV}}$. Only those data satisfying $A_{\text{\upshape FOV}}>0.75$, $106^\circ < \phi < 164^\circ$, and $-40^\circ < \theta < 40^\circ$, are used to reduce the possible impact of FOV effects \citep[see also][]{nir26p04,yog26p25}. Here $\phi$ and $\theta$ are azimuth and elevation with differential energy flux having the maximum measured value.  Third, the velocity is required in the range $200-600$ km/s, indicating the usual cases with moderate velocity. Fourth, to reduce the mixing effect of radial evolution of solar wind turbulence, we focus on the narrow region with heliocentric distances between 0.1 and 0.2 au. In addition, ion-scale waves are well detected merely when the solar wind velocity is quasi-parallel (or antiparallel)  to the background magnetic field. The angle between ${\bm V}_p$ and ${\bm B}$, $\theta_{vb}$, is required in the range of $\theta_{vb} < 45^\circ$, where the angle $\theta_{vb}$ is restricted to be between $0^\circ$ and $90^\circ$ for convenience.

\begin{figure}
\epsscale{1.2} \plotone{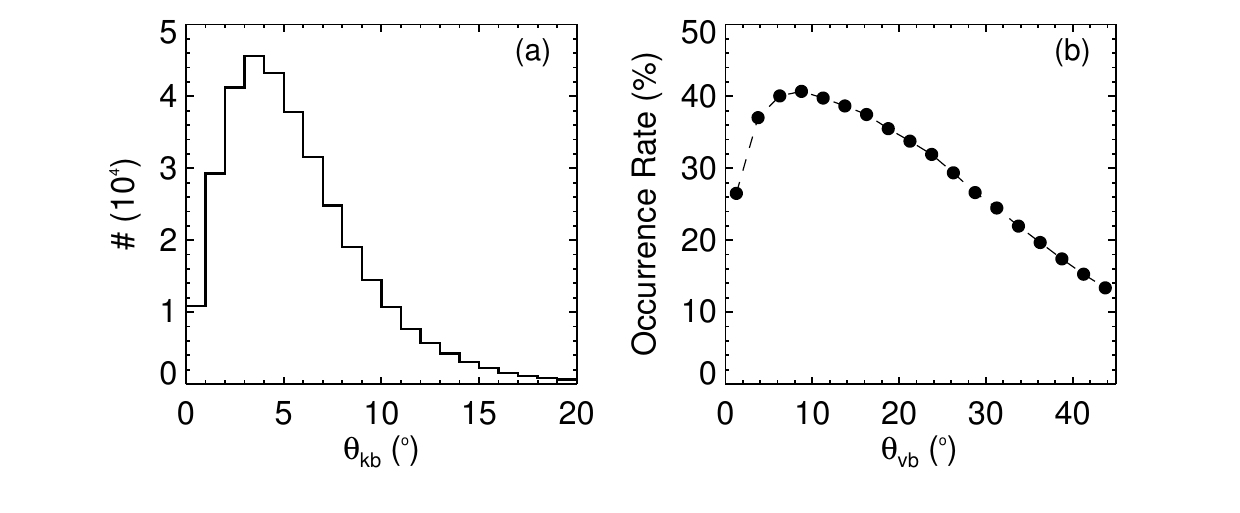}
\caption{Panel (a): distribution of wave propagation angle ($\theta_{kb}$). Panel (b): wave occurrence rate against the angle between ${\bm V}_p$ and ${\bm B}$ ($\theta_{vb}$).}
\label{Fig1}
\end{figure}

\begin{figure}
\epsscale{1.2} \plotone{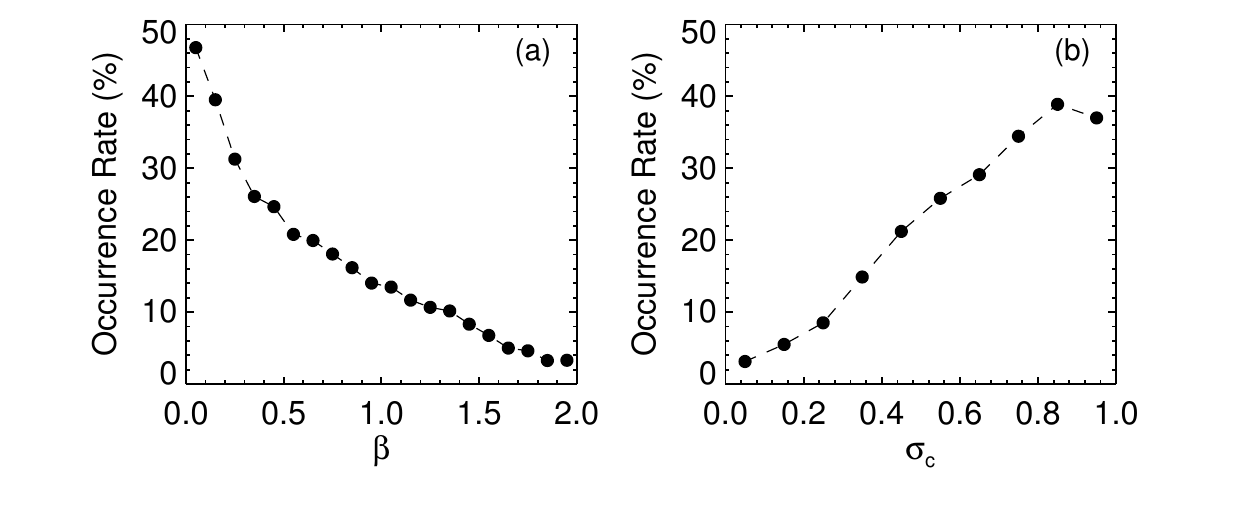}
\caption{Panel (a): wave occurrence rate against plasma beta ($\beta$). Panel (b): wave occurrence rate against cross helicity ($\sigma_{c}$).}
\label{Fig2}
\end{figure}

Two primary steps are conducted to detect waves, following the method developed in our previous studies \citep{zha17p08,zha18p15}. For a given time interval of magnetic field, the first step is to calculate the reduced magnetic helicity spectrum in a field-aligned coordinate system (with the $z$ direction along the ambient magnetic field). The magnetic helicity spectrum is normalized and has values in the range from $-1$ to 1 \citep{mat82p11,hej11p85}. The spectrum values are examined for the frequency around the local proton cyclotron frequency. If the spectrum has absolute values $\geqslant$ 0.7 in some frequency band with a minimum bandwidth of 0.5 Hz, the second step is carried out to identify an enhanced power spectrum. The enhancement here means that transverse power with large spectrum values of magnetic helicity is three times larger than the background power in the same frequency band, where the background power is determined by a power law fit for the entire transverse power spectrum. A wave is recorded if the above two steps are fulfilled. To well capture the small scale waves, each time period of 10 minutes is divided into short and overlapping time intervals, and each short time interval is set to be 10 s; the overlap time is set to be 5 s to increase the sample data. After these steps, the time intervals with wave occurrence can be collected and the wave occurrence rate can be obtained. The left-handed and right-handed polarization senses of the waves are determined by the sign of spectrum values of magnetic helicity; the minus corresponds to the left-handed polarization in our study. The wave propagation angle with respect to background magnetic field, $\theta_{kb}$, is estimated by minimum variance analysis, where the angle $\theta_{kb}$ is also restricted to be between $0^\circ$ and $90^\circ$.
\begin{figure}
\epsscale{0.7} \plotone{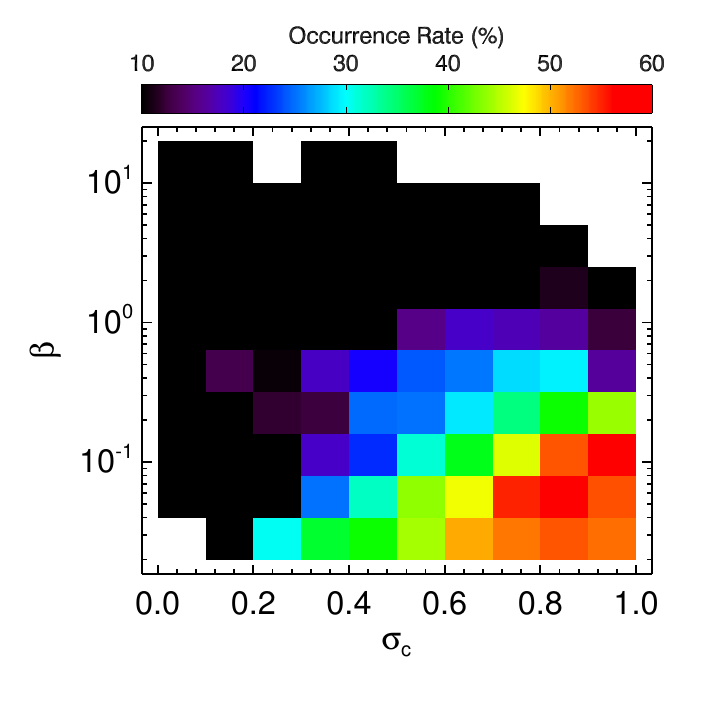}
\caption{Color scale plot of wave occurrence rate in the ($\sigma_c$, $\beta$) space, where cells with sample number less than 10 are indicated by the white color}.
\label{Fig3}
\end{figure}

\begin{figure}
\epsscale{0.7} \plotone{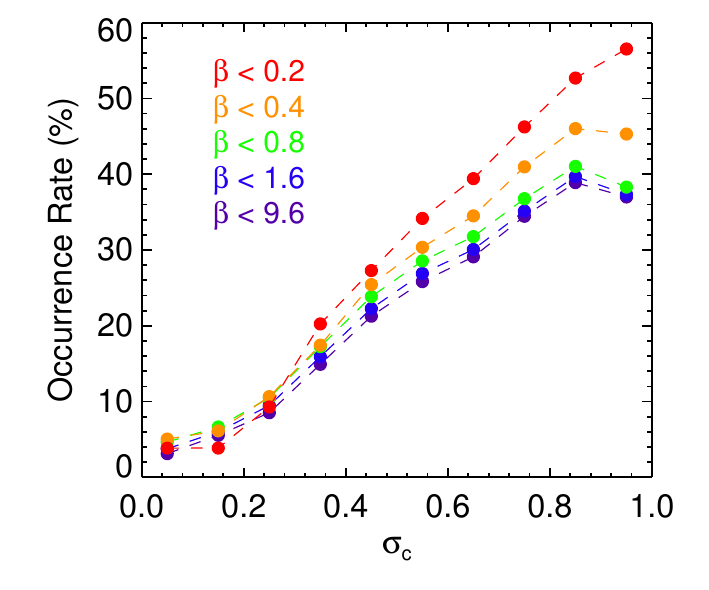}
\caption{Wave occurrence rate against cross helicity ($\sigma_{c}$), where $\beta < 0.2$ is used.}
\label{Fig4}
\end{figure}

\newpage
\section{Statistical Results}  %
In total $\sim1.1 \times 10^6$ time intervals are examined, and $\sim4.3 \times 10^5$ intervals are recorded as in the presence of ion-scale waves, corresponding to a total occurrence rate $\sim37.8\%$. Among the intervals associated with waves, the majority (78.9$\%$) of intervals are shown with left-handed waves. This Letter focuses on these left-handed waves, which are usually believed to be ICWs \citep{jia09p05,bow20p74,liu23p69}. Figure \ref{Fig1} presents an overview of the distribution of wave propagation angle (panel (a)) and the occurrence rate of left-handed waves against the angle between ${\bm V}_p$ and ${\bm B}$ (panel (b)). It is shown that these waves are mainly of approximately parallel propagation and most of them are detected when ${\bm V}_p$ is quasi-parallel to ${\bm B}$;  $88.4\%$ of cases have propagation angle less than $10^\circ$, and their occurrence rate peaks at $\theta_{vb} \simeq 8.8^\circ$. These properties are consistent with existing results \citep{jia09p05,jia10p15,liu23p69}.

Figure \ref{Fig2} shows the main findings of this Letter. The first finding is the dependence of the wave occurrence rate on $\beta$. Panel (a) plots the occurrence rate with respect to $\beta$. The occurrence rate drops from 46.8$\%$ to 19.9$\%$ with $\beta$ changing from 0.05 to 0.65, and it reduces further with $\beta$ increasing. The second finding is the dependence of the wave occurrence rate on  $\sigma_c$ (panel (b)). One can see that the occurrence rate increases with  $\sigma_c$ in principle. The increase becomes particularly fast when $\sigma_c$ exceeds 0.3, except for the odd case with $\sigma_c$ approaching unity. We will show later that this odd case disappears for $\beta<0.2$. 

Figure \ref{Fig3} displays a color scale plot of the occurrence rate in the ($\sigma_c$, $\beta$) space, where the red color marks an occurrence rate $60\%$ and cells with sample number less than 10 are indicated by the white color. It is interesting that the highest occurrence rate ($> 50\%$) precisely appears in bottom right corner of the panel, where $\beta$ is low and meanwhile $\sigma_c$ is high. In the opposite corner, however, the occurrence rate is generally low ($< 10\%$), indicated by the black color. 

From Figure \ref{Fig3}, one may realize that the low $\beta$ is not a unique factor to result in a high occurrence rate of ion-scale waves. The combined effect of $\beta$ and $\sigma_c$ could play a critical role in the occurrence of the waves. The occurrence rate is still low if $\sigma_c$ is not high, even though $\beta$ is low. Figure \ref{Fig4} plots the occurrence rate against $\sigma_c$, in which the lines with different colors correspond to different $\beta$ ranges. one can see that all color lines nearly overlap and the values of occurrence rate are smaller than $\sim10\%$ when $\sigma_c < 0.3$. When $\sigma_c > 0.3$, the occurrence rate increases rapidly with $\sigma_c$ and becomes sensitive to $\beta$ ranges for $\beta < 0.8$. For the case $\beta < 0.2$, the occurrence rate is significantly high (up to 56.5$\%$) with $\sigma_c$ approaching 0.95, and exceeds 45$\%$ as $\sigma_c$ is larger than 0.7. 

\section{Summary and Discussion}
Based on in situ observations of ion-scale waves and plasma parameters in the near-Sun solar wind, this Letter investigates the occurrence rate of left-handed polarized waves in ($\sigma_c$, $\beta$) space for the first time. It is found that the occurrence rate significantly depends on not only $\beta$ but also $\sigma_c$. It decreases as $\beta$ increases, and increases as $\sigma_c$ increases. It is generally low when $\sigma_c$ is low ($< 0.3$). When $\sigma_c$ becomes higher than 0.3, the occurrence rate rises rapidly with $\sigma_c$. Consequently, the high occurrence rate ($\sim 50\%$) occurs as $\sigma_c$ is high enough ($\sigma_c > 0.7$) and meanwhile the low-$\beta$ condition ($\beta < 0.2$) is fulfilled.  

In the present study $\sigma_c$ is an inertial-scale parameter to describe the imbalance of turbulence while wave occurrence rate is an ion-scale quantity to assess the wave occurrence. The clear dependence of the wave occurrence rate on $\sigma_c$ suggests that inertial-scale turbulence likely controls ion-scale wave occurrence. Such cross-scale control becomes significantly strong when $\beta$ is low, implying the crucial role of plasma $\beta$  in affecting the cross-scale control from the cross helicity.       

The present observations provide support for the helicity barrier mechanism \citep{mey21p01,squ22p15}. Theoretically, this mechanism contributes to the generation of ICWs during the process of evolution of imbalanced magnetized turbulence. For highly imbalanced turbulence with high $\sigma_c$, the turbulent cascade cannot proceed to pass the ion gyroradius scale through the standard critically balanced cascade. The helicity barrier mechanism predicts that energy accumulates above ion scales. As demonstrated by \citet{squ22p15}, this accumulated energy drives a parallel cascade that ultimately generates ICWs. This mechanism works efficiently merely under the low-$\beta$ condition. The barrier weakens and eventually disappears as $\beta$ increases (roughly $\beta \gtrsim 1$), because the ion gyroradius scale and the scale where the barrier operates become separated and compressive channels open up. The observations that the highest occurrence rate ($> 50\%$) precisely appears in the low-$\beta$, high-$\sigma_c$ regime are thus well consistent with the theory of the helicity barrier. 

Particularly striking in the present observations is that the highest wave occurrence rate is confined to the low-$\beta$, high-$\sigma_c$ corner of parameter space (Figure \ref{Fig3}), indicating that both conditions are required simultaneously, exactly as predicted by the helicity barrier theory. Moreover, unlike existing observational signatures associated with kinetic turbulence, the simultaneous dependence of ion-scale wave occurrence on $\beta$ and $\sigma_c$ is a direct prediction of the helicity barrier theory \citep{mey21p01,squ22p15}.

The present discussion may be helpful to provide insight into the solar coronal heating problem, in which ion cyclotron wave version has been suggested for the heating \citep{cra02p29,hol02p47,ise11p88}. The difficulty of this version is that it is unclear how ICWs are generated in the corona \citep{how00p73,she83p25}. We propose a scenario for the ICW generation and subsequent heating of the corona as follows. Large-scale Alfv\'en waves are theoretically and observationally ubiquitous in the transition region (below the corona) and the corona \citep{tom07p92,hej09p25,mci11p77,che22p37}. The Alfv\'en waves are outward propagating and part of them suffer from wave reflection as Alfv\'en speed increases with height, which results in counter-propagating waves. The counter-propagating waves initiate chaotic nonlinear interactions and drive turbulence that transfers energy progressively from large scales to smaller scales. The driven turbulence is highly imbalanced with high $\sigma_c$. Under the condition of low $\beta$, which is generally fulfilled in the solar inner corona \citep{gar01p71}, the helicity barrier mechanism works efficiently and contributes to the generation of ICWs. The cyclotron waves heat the corona as well as the solar wind through cyclotron resonant wave-particle interactions when they propagate away from the Sun \citep{kas08p03,zha20p14,bow22p01}. We remark that this scenario is preliminary and a detailed understanding of the scenario is desirable in future research.

Note that a recent study by \citet{nir26p04} also used PSP data from Encounters 3–24 and conducted a mission-wide statistical analysis on ion-scale waves, covering radial distances from about 0.05 to 0.28 au. The prevalence of left-handed waves, their parallel character, and their strong dependence on the angle $\theta_{vb}$, have been found in \citet{nir26p04}. The authors showed that left-handed waves are frequently observed than right-handed waves, with the time fraction of left-handed wave occurrence reaching $\sim30\%$ within 0.15 au. The left-handed waves in their observations are preferentially observed when $\theta_{vb}$ is smaller (or larger close to $180^\circ$), exhibiting strong dependence on $\theta_{vb}$. The authors thus argued that the observed left-handed waves are parallel propagating.  Our observations (Figure \ref{Fig1}) are consistent with their results; $88.5\%$ of the waves have propagation angle less than $10^\circ$ and $65.8\%$ are detected with $\theta_{vb} < 25^\circ$ in our observations. In particular, we focus on the correlations of wave occurrence with $\beta$ and $\sigma_c$ and reveal the simultaneous dependence of the wave occurrence on $\beta$ and $\sigma_c$, which is the genuinely new result of our study. The methods of detecting waves in the both studies are different. \citet{nir26p04} divided the long time period of observations into continuous 15 minute intervals and performed Morlet wavelet transforms of magnetic field fluctuations. Ion-scale waves were identified by selecting the frequency–time regions with high values of circular polarization and were stored with a time cadence of 1 NYs ($\sim$0.874 s). 
In our method we divide the long time period into overlapping 10 s intervals and conduct Fourier transforms of the fluctuations. Ion-scale waves are recorded with magnetic helicity threshold 0.7, a minimum bandwidth 0.5 Hz, and an enhanced-power feature. The overlap time is set to be 5 s. Consequently the time cadence of our wave record is 5 s. We also note that the occurrence rate of left-handed waves in our study decreases from $39.8\%$ to $18.0\%$ as radial distances increase from 0.1 to 0.2 au while that of right-handed waves increases from $3.4\%$ to $12.4\%$ with distances from 0.1 to 0.13 au, and becomes approximately a constant  around 8\% until 0.2 au (not shown), which share similar trends with the time fractions of wave occurrence in \citet{nir26p04} in the same range of radial distances. 

\acknowledgments
The authors acknowledge the NASA Parker Solar Probe Mission. All data used in this Letter are publicly available via the website (https://cdaweb.gsfc.nasa.gov/pub/data/). This research was supported by the National Natural Science Foundation of China (No. 42474211) , the Program for Innovative Research Team (in Science and Technology) in University of Henan Province (No. 25IRTSTHN014), Scientific and Technological Innovation Talents in University of Henan Province (No. 24HASTIT033), and partly by the Key Scientific Research Project in University of Henan Province (No. 25B170001). R.M. acknowledges support from the U.S. Department of Energy under grant DE-SC0026201. The authors thank the anonymous referee for valuable comments that improved this Letter.


\end{document}